# Quantification of the bond-angle dispersion by Raman spectroscopy and the strain energy of amorphous silicon

P. Roura[#], J. Farjas[#] and P. Roca i Cabarrocas[&]

[#]GRMT, Department of Physics, University of Girona, Campus Montilivi, E17071-Girona, Catalonia, Spain. pere.roura@udg.es, jordi.farjas@udg.es
Tel: (34) 972 418383, Fax: (34) 972 418098
[&]LPICM, Ecole Polytechnique, 91128 Palaiseau, France. pere.roca@polytechnique.edu
Tel: (33) 1 69334314 Fax: (33) 1 69334333

A thorough critical analysis of the theoretical relationships between the bond-angle dispersion in a-Si, $\Delta\theta$, and the width of the transverse optical (TO) Raman peak, $\Gamma$, is presented. It is shown that the discrepancies between them are drastically reduced when unified definitions for $\Delta\theta$ and $\Gamma$ are used. This reduced dispersion in the predicted values of $\Delta\theta$ together with the broad agreement with the scarce direct determinations of $\Delta\theta$ is then used to analyze the strain energy in partially relaxed pure a-Si. It is concluded that defect annihilation does not contribute appreciably to reducing the a-Si energy during structural relaxation. In contrast, it can account for half of the crystallization energy, which can be as low as 7 kJ/mol in defect-free a-Si.

**I.- Introduction**

Structural disorder in pure amorphous silicon (a-Si) can be modelled by a defect-free continuous random network of silicon atoms with tetrahedral coordination [1]. This kind of model is currently considered to be an approximation to the real material structure [2]. In fact, a continuous network is achieved at the expense of highly distorted bond angles, $\theta$. This built-in strain results in an intrinsic structural instability. The highly strained bonds tend to break and defects, notably dangling bonds and, perhaps, floating bonds, are created. This evolution has been modelled [2] and explains the high density of dangling bonds in pure a-Si, which makes this material useless for electronic applications.

When hydrogen is added, it saturates the dangling bonds and relaxes the Si network stress [3]. These beneficial effects render hydrogenated amorphous silicon (a-Si:H) a technologically important material because its optical and electrical properties are improved. Although, when compared with pure a-Si, the covalent Si network in a-Si:H is considered to be "relaxed", it is known that the degree of relaxation depends on the deposition conditions [4]. Consequently, it is worthwhile developing accurate experimental tools for quantifying the bond angle dispersion, $\Delta\theta$.

Unfortunately, direct measurements of $\Delta\theta$ by X-ray [5, 6] or neutron [7] diffraction involve very delicate experiments and analyses. However, indirect, routine quantification of bond angle disorder is carried out using Raman spectroscopy. Experiments [7] and theory [8] show that the transverse optical (TO) band broadens as $\Delta\theta$ increases. The problem is that, in the literature, a number of theoretical relationships between $\Delta\theta$ and the transverse optical bandwidth, $\Gamma$,

$$\Delta\theta = f(\Gamma) \qquad (1)$$

have been proposed which show notorious discrepancies between them. The lack of solid experimental confirmation makes using one particular relationship for quantifying $\Delta\theta$ a question of personal choice.

In this paper, we present a critical review of the theoretical results published so far and compare them with the (scarce) direct measurements of $\Delta\theta$. This analysis provides reasonable arguments in favour of one particular relationship. In a second stage, this relationship will be applied to analyzing the amount of heat that evolves from pure a-Si when it is relaxed by thermal annealing.

## II.- Literature review and critical analysis of the $\Delta\theta = f(\Gamma)$ relationship

### II.1 Theoretical results

In Fig. 1, we have plotted all the $\Delta\theta = f(\Gamma)$ relationships so far published. Except for the one by Tsu et al. [8], all of them have been obtained by numerical simulation of the Raman spectra on a series of microscopic models of a-Si. Beeman et al. [9] made the first calculations with the aim of confirming the validity of Tsu's relationship:

$$\Gamma^2 = 32^2 + (6.75\Delta\theta)^2, \qquad (2)$$

where $\Gamma$ (in cm$^{-1}$) is the full width at half maximum (FWHM) of the TO band and $\Delta\theta$ (in degrees) is the root mean square (RMS) deviation with respect to the tetrahedral angle of the bond angle distribution. For one particular choice of the Si-Si bond polarizability, they obtained a result that was very close to that of Tsu and proposed a formula that, in fact, was just the linearization of Tsu's formula in the $\Delta\theta$ range of practical interest:

$$(\Gamma/2) = 7.5 + 3\Delta\theta, \qquad (3)$$

where, now, $(\Gamma/2)$ is the half width at the high energy side of the TO band (see Fig.2). Eq. (3), which herein will be referred to as the Beeman-Tsu formula, is the one used most in the literature. For this reason, the rest of the results plotted in Fig. 1 take the work by Beeman et al. [9] as reference.

The reduced Raman spectrum, $I(\omega)$, with the thermal and harmonic oscillator factors removed, can be expressed as the product of the vibrational density of states, $g(\omega)$, and a coupling parameter $C(\omega)$:

$$I(\omega) = C(\omega) g(\omega) . \qquad (4)$$

$g(\omega)$ depends on the microscopic model of a-Si, whereas $C(\omega)$ is a function of the bond polarizability. In their simulations, Vink et al. [10] used larger and more realistic a-Si models than those of Beeman, leading to presumably more accurate densities of states. However, in view of their results, it is clear that the main difference was the choice of the coupling constant. They were led to the conclusion that the best results were obtained with a $C(\omega)$ function obtained experimentally which is almost constant in the range of the TO band frequencies. The result of this particular choice has been plotted in Fig. 1, where we see that the discrepancy with the Beeman-Tsu curve is significant.

The calculations carried out by Wong et al. [11] delivered the density of vibrational states for two microscopic models of a-Si. Consequently, their results are equivalent to considering a constant coupling parameter. The original values of Wong et al. [11] have been corrected because the FWHM was given instead of $(\Gamma/2)$ (see Fig. 2). The conversion from FWHM/2 to $(\Gamma/2)$ has been obtained from the inset of Fig. 2. In this inset, we have plotted the widths of the TO bands published by Battaglia et al. [12] corresponding to pure a-Si. These spectra have an outstanding signal to noise ratio and cover the largest range of $(\Gamma/2)$ ever measured in a single study. In addition, we obtained the points of hydrogenated films. In both cases the measured spectrum, $I_{meas}$, was converted into the reduced spectrum according to:

$$I(\omega) = \frac{I_{meas}(\omega)\omega}{n(\omega,T)+1}, \tag{5}$$

where $n(\omega,T)$ is the number of phonons with frequency $\omega$ at temperature T. According to the inset of Fig. 2 a linear relationship between the FWHM/2 and $(\Gamma/2)$ can be assumed, FWHM/2 = -15+1.5$(\Gamma/2)$. After this correction, the results of Wong et al. [11] still depart from the Beeman-Tsu relationship (Fig. 1). In this figure, we realize that Vink's relationship predicts similar values of $(\Gamma/2)$ for the range of the bond angle dispersion modelled by Wong et al. This is not surprising, as the coupling parameter is almost constant in Vink's calculation as well [10].

From Fig. 2, it is worth noting that $(\Gamma/2)$ is very similar in the measured and reduced spectra and that the FWHM is clearly larger in the measured spectrum. This fact and the dependence of the low-energy side of the TO band on the medium range order [13] makes it preferable for quantifying the TO width with $(\Gamma/2)$. Unfortunately, the results of many papers cannot be accurately analyzed because it is not clear which is the parameter used to characterize the TO bandwidth.

The curve of Maley et al. [14] cannot be directly compared with eq. (3). This is due to two reasons. First, as Wong et al. did in ref. [10], Maley et al. calculated the FWHM of the reduced Raman spectrum and not its half width at high energy, $(\Gamma/2)$. Second, they carried out their simulations with a-Si models whose bond angle distributions were not Gaussian-like and decided to quantify their dispersion through the width of the Gaussian distribution, $\Delta\theta_G$, that best fitted the actual distribution. In other words, the curve labelled "Maley" in Fig. 1 corresponds to the relationship between the FWHM/2 and $\Delta\theta_G$. From the values of $\Delta\theta_G$ and $\Delta\theta$ detailed in their paper [14], we have deduced the linear relationship $\Delta\theta_G = 1.59 + 0.77\Delta\theta$. When the FWHM and $\Delta\theta_G$ are converted into $(\Gamma/2)$ and $\Delta\theta$ respectively, the corrected relationship of Maley is very close to that of Beeman-Tsu. This coincidence is very reassuring because, although the microscopic models used by Maley et al. [14] were similar to those of Beeman et al. [9], different polarizability mechanisms were assumed.

Finally, a model calculation without adjustable parameters by Marinov et al. [15] gave $(\Gamma/2) = 40$ cm$^{-1}$ (FWHM = 90 cm$^{-1}$) for a model with $\Delta\theta = 10.8°$, in exact agreement with the Beeman-Tsu formula (eq. (3)).

II.2 Comparison with experiment

As we have already mentioned in the introduction, direct measurements of $\Delta\theta$ using diffraction techniques are very scarce. In fact, we have only found three papers [5, 7, 16] reporting measurements of the radial distribution function (RDF) on amorphous silicon materials that are of suitable quality for our discussion.

The first diffraction experiments were carried out by Schülke on a-Si:H [16]. He obtained a value of $\Delta\theta = 7.9 \pm 0.4°$. Although no Raman measurements were reported, we know that, since a-Si:H has a relaxed Si network, for this material the range of $(\Gamma/2)$ is quite narrow. Our own measurements of a-Si:H samples without substrate indicate that a good estimation is $33 \pm 3$ cm$^{-1}$. Although wider Raman spectra have been reported in the literature, we should keep in mind that macroscopic stress due to the substrate [17] as well as film inhomogeneity [18] tend to increase $(\Gamma/2)$. Our estimation allows us to plot the result of Schulke in Fig. 1 (diamond).

The original relationship of Maley was tested by Fortner et al. [7] with the RDF determined by neutron diffraction experiments carried out at room temperature on films prepared by RF sputtering. The experimental points (open circles) show reasonable agreement with Maley's curve. If we correct the FWHM/2 and $\Delta\theta_G$ values of the experimental points as we did for the theoretical curve, this agreement would be translated to the Beeman-Tsu curve. However, analyzing the experimental RDF [7] allows direct evaluation of $\Delta\theta$. A good fit to the second RDF peak is obtained with a Gaussian distribution with $\Delta\theta_G = 9.9°$ and 11.5° centered around $\bar{\theta} = 108.4°$ and 108.6° respectively. From these values calculating $\Delta\theta$ is straightforward ($\Delta\theta = [(\Delta\theta_G)^2 + (\theta_t - \bar{\theta})^2]^{1/2}$ where $\theta_t = 109.47°$ is the tetrahedral bond angle) and the $\Delta\theta$ values are almost identical to $\Delta\theta_G$ ($\Delta\theta_G$ and $\Delta\theta$ are also similar for the RDFs measured by Laaziri et al. [5, 6] commented below). Our analysis agrees with the claim [19] that large deviations of bond angle distributions from Gaussian are due to the finite size of the microscopic models. Consequently, we should consider that the standard deviation of the Gaussian distribution used to fit the experimental RDF is nearly the RMS of the actual distribution. We have, thus, corrected only the Raman width of the points of Fortner et al. [7] but not the angle dispersion. The result is plotted as solid circles in Fig. 1.

Finally, the most precise diffraction experiments have recently been reported by Laaziri et al. [5, 6]. They carried out x-ray diffraction on two impurity-free, dense a-Si samples, at a low temperature to reduce thermal broadening effects on the RDF. For all these reasons, the results of Laaziri et al. [5, 6] constitute a reference for the actual microscopic models of a-Si [2, 19, 20, 21]. Unfortunately, the Raman spectra of the same samples were of low quality and ($\Gamma/2$) values have large error bars [22]. The results are plotted in Fig. 1 as solid squares.

The wide dispersion of the experimental points in Fig. 1 does not allow choosing one of the theoretical predictions as the correct one. The ensemble of points are closer to the Beeman-Tsu relationship, consequently we think that, if one single relationship exists between ($\Gamma/2$) and $\Delta\theta$ (i.e. a relationship that is independent of the particular microstructure), it is closer to that of Beeman-Tsu than to that of Vink. This conclusion is reinforced by the fact that four of the six theoretical calculations discussed in the previous subsection deliver values of $(\Gamma/2)$ that are very close to eq. (3). In view of the coincident slopes of Vink's and Beeman-Tsu's formulas (Fig. 1), probably the correct relationship will have the general form: $(\Gamma/2) = A + 3\Delta\theta$ with $12 < A < 7.5$ cm$^{-1}$. For the rest of the paper, we will use eq. (3) for the analysis of the relaxation energy of a-Si given below. However, the quantitative results obtained here would change only slightly if Vink's relationship was used instead of that of Beeman-Tsu.

**III. Microscopic interpretation of the energy of relaxation and crystallization of pure a-Si**

When pure a-Si is heated, its structure evolves irreversibly towards states of lower energy through structural relaxation and crystallization processes [23]. Experiments have shown that, during relaxation, the bond angle dispersion and the density of point defects diminish [5, 8, 12, 24, 25]. This evolution has been simulated by molecular dynamics, too [26, 27]. However, the relative contribution of both effects to the relaxation energy has been the object of controversy for many years. Whereas Stolk et al. [24] assign most of the relaxation energy to bond-angle strain, Roorda et al. [25, 28] assign it to defect recombination. In contrast, little attention has been paid to the crystallization energy. It seems that there is ample agreement that it can be almost entirely explained as arising from strain [29, 30].

In this section we will apply the Beeman-Tsu formula (eq.(3)) for quantifying the strain energy stored in a-Si by reanalyzing the excellent calorimetric experiments carried out by Roorda et al. [25] concerning the structural relaxation and crystallization of a-Si obtained by ion-implantation. In these experiments, a number of identical samples were partially relaxed by annealing them at different temperatures (from 300 to 773 K) for 45 min. The degree of relaxation reached in each sample was then analyzed by differential scanning calorimetry. Every sample was heated in a calorimeter at 40 K/min from room temperature until a maximum temperature where complete crystallization was reached. The total heat evolved during the process (i.e., the evolution from a-Si to c-Si), $Q_{ac}$, was recorded. From the typical thermogram shown in the inset of Fig. 3 it is clear that $Q_{ac}$ can be understood as the addition of two components:

$$Q_{ac} = Q_{relax} + Q_{cryst}. \tag{6}$$

The heat of relaxation, $Q_{relax}$, evolves at a low temperature and gives an unstructured band in the thermogram whereas the heat of crystallization, $Q_{cryst}$, is released at high temperatures and forms a sharp peak. For the samples annealed at a high temperature, $Q_{relax}$ was lower whereas $Q_{cryst}$ remained constant, independent of the annealing temperature. For the sample annealed at 773 K, $Q_{relax} \approx 0$ and $Q_{ac} = Q_{cryst} = 13.7$ kJ/mol. We can, thus, decompose $Q_{ac}$ into its components for any annealing temperature as sketched in Fig. 3.

In order to understand the relationship between $Q_{ac}$ and the energy stored as strain in a-Si, we should realize that the vibrational energy (the zero point energy and heat capacity terms) contribute with less than 10% to the value of $Q_{ac}$. In other words, the heat evolved $Q_{ac}$ can be considered as the total static energy of a-Si at a given stage of relaxation, $U_{ac}$ (see Appendix). From the microscopic point of view, we can consider that the energy of a-Si has two terms. One is due to bond strain and the other, to structural defects:

$$Q_{ac} \approx U_{ac} = U_{strain} + U_{defects}. \tag{7}$$

The strain energy of a-Si arises from bond stretching (variations of the bond length r) and from bond bending (variations of bond angle $\theta$), and it can be written as [29, 31]:

$$U_{strain} = N2\beta r^2 \left[1 + \frac{3}{2}\frac{\alpha}{\beta}\left(\frac{\Delta r}{r}\right)^2 \frac{1}{(\Delta\theta)^2}\right](\Delta\theta)^2 \approx N2\beta r^2 (\Delta\theta)^2, \qquad (8)$$

Where N is the number of atoms in the material, $\Delta r$ its standard deviation with respect to the value in c-Si and $\alpha$ and $\beta$ are the bond-stretching and bond-bending force constants of the Keating potential [32] respectively. EXAFS experiments done in a-Si and a-Si:H materials have revealed that $\Delta r$ is proportional to $\Delta\theta$ [31]: $(\Delta r / r\Delta\theta)^2 = 8.3\ 10^{-3}$. A lower value ($5.6\ 10^{-3}$) is estimated from XRD experiments [6]. On the other hand, by fitting the Raman spectra and the density of vibrational states of a-Si to atomistic models, several authors have obtained $\alpha/\beta$ values in the 2.9 to 6.7 range [33, 34]. With these values, the expression inside the brackets departs from unity by less than 8%. This means that, as indicated in the right-hand side of eq. (8), the most important contribution to $U_{strain}$ is bond bending.

From the experiments by Roorda et al. [25] we can obtain the experimental dependence between $Q_{ac}$ and $(\Delta\theta)^2$ in partially relaxed samples. $Q_{ac}$ is delivered by calorimetry whereas Raman measurements made on the same samples allow $\Delta\theta$ to be quantified by applying the Beeman-Tsu formula. The squares in Fig. 3 correspond to the values of $\Delta\theta$ and $Q_{ac}$ measured by Roorda et al. [25] on the same samples, whereas stars are obtained by combining the Raman results determined by Battaglia et al. [12] and the calorimetric experiments of Mercure et al. [28]. As we commented above, the series of points does not continue until $(\Delta\theta)^2 = 0$ because crystallization impedes further relaxation.

III.1 Structural relaxation

From Fig. 3, we see that $U_{ac}$ is not proportional to $(\Delta\theta)^2$ but that the points keep a linear relationship with an extrapolated positive value of $U_{ac}$ at $(\Delta\theta)^2 = 0$. The most direct interpretation of this result is that, in addition to strain, the contribution of defects to $U_{ac}$ is not negligible. Therefore, with the help of eq. (8) we can rewrite eq. (7) as:

$$U_{ac} = K_{\Delta\theta} \cdot (\Delta\theta)^2 + U_{defects}, \qquad (9)$$

where $K_{\Delta\theta}$ is detailed in eq. (8). At this point it is necessary to accurately define the meaning of $U_{defects}$. As discussed extensively by Roorda et al. [25], a structural defect may produce around it a local distortion of the covalent Si-Si network. So, a defect population may increase $\Delta\theta$ of the material. Consequently, in addition to the 'chemical' energy, $U_{defects}$, associated with the bonding (for instance dangling-bonds or floating bonds), the defects may increase the overall strain energy [the first term of eq. (9)].

From the nice linear behavior shown in Fig. 3, it is tempting to consider that $U_{defects}$ is constant throughout the relaxation process. However, all experiments [12, 24, 35] indicate that the defect density diminishes during the structural relaxation of pure a-Si. Thus, we conclude that $U_{defects}$ must diminish as the annealing temperature increases and, consequently, that the slope of the experimental points must be an upper bound to $K_{\Delta\theta}$. This conclusion is translated through eq. (8) to $\beta$ whose value should be lower than $4.1 \pm 0.4$ N/m. Our estimate is lower than the values obtained from simulating Raman spectra ($6.7 < \beta < 16.6$ N/m [33, 34]). For this reason, we think that the strain energy must be close to the dashed line of Fig. 3, otherwise the discrepancy with the published values of $\beta$ would increase. In fact, the value of $U_{ac}$ is more sensitive than the shape of the Raman spectra to variations of $\beta$. The analysis of Marinov et al. [15] shows that the most prominent feature in the Raman spectrum of a-Si (the TO band) has a stretching character and, consequently, it has a minor dependence on the bending force constant $\beta$.

Since the slope must be close to that of the dashed line of Fig. 3, we conclude that the (chemical) energy associated with defects, $U_{defects}$, remains almost unchanged during structural relaxation (i.e. $|\Delta U_{defects}| << U_{defects}$ where $|\Delta U_{defects}|$ means the total variation of the defect energy between the as-implanted and relaxed states). Therefore, its contribution to the heat of relaxation, $Q_{relax}$, must be much lower than that of strain. Furthermore, from the experiments of Stolke et al. [24], we can also conclude that defects do not contribute appreciably to the strain energy, $U_{strain}$. This is so because these experiments have shown that, for samples relaxed to a given intermediate value of $\Delta\theta$, the density of defects depends on how this state has been reached [24]. For a given value of $\Delta\theta$ the density of defects can vary by as much as a factor of 3, so $\Delta\theta$ is quite independent from the density of defects.

Our conclusion that defect annihilation does not contribute significantly to the heat of relaxation poses severe restrictions on the nature of the defects that recombine during structural relaxation. For instance, from diffusivity [36] and carrier lifetime experiments [24], we know that the amount of defects that recombine during structural relaxation taking place between 300 and 773 K is around 0.5-1% atomic. From our Fig. 3, an upper bound of 1 kJ/mol for the contribution of defects to $Q_{relax}$ can be assumed (this contribution would imply decreasing $\beta$ to below 3.5 N/m, i.e. to ½ of the theoretical estimates). This results in a formation energy per defect lower than 2-1 eV. Thus, the defects that recombine during relaxation can hardly be vacancies (formation energy of 4 eV [37]).

III.2 Crystallization

We believe that the most striking conclusion to be drawn from Fig. 3 is the interpretation of the heat of crystallization. Our analysis leads us to conclude that, just before crystallization (i.e. the first point in Fig. 3), about one-half of the a-Si energy is associated with structural defects. This is in contrast with the early interpretation of the value of $Q_{cryst}$ [29, 30] which considered that most of the energy of relaxed a-Si was stored as bond-angle strain, i.e.:

$$Q_{cryst} \approx U_{cryst} \approx K \cdot (\Delta\theta)^2. \tag{10}$$

As far as we know, this interpretation is widely accepted without suspicion. However, it is false because the value of K that fits the first point of Fig. 3 (sample relaxed at 77 K) delivers a strain energy value that is clearly higher than the experimental energy for the partially relaxed states (this situation is sketched in Fig. 4).

An interesting prediction of our analysis is that the heat of crystallization of a-Si may be lower than the reported values of 11-14 kJ/mol [23, 38, 39]. If the density of defects were lower, $Q_{cryst}$ could be as low as 6.6 kJ/mol. In fact, the first crystallization experiments of a-Si reported a value of 9.2 kJ/mol [40]. This low value was subsequently considered to be erroneous by other authors.

III.3 Earlier analyses

Our analysis of $Q_{ac}$ based on eq. (9) agrees with that carried out by Stolk et al. [24]. However, the conclusion of these authors concerning the contribution of defects to $Q_{relax}$ was very ambiguous. On the other hand, we think that our eq. (9) is more realistic than the analysis given by Roorda et al. [25] who supposed that all the heat released during relaxation ($Q_{relax}$) was due exclusively to defect annihilation:

$$Q_{relax} \propto \delta N_d \propto \delta(\Delta\theta) \tag{11}$$

where $\delta N_d$ is the reduction of defect concentration during relaxation. Additionally, they considered that this reduction produced a diminution of $\Delta\theta$ proportional to $\delta N_d$. Consequently, a proportionality between $Q_{ac}$ and $\delta(\Delta\theta)$ was predicted. Despite of the reasonable alignment of the experimental points in Fig. 4, we think that this interpretation of $Q_{relax}$ is not correct. First, if $\Delta\theta$ increased with the defect concentration, then the associated strain energy would follow the $(\Delta\theta)^2$ dependence (the parabola of Fig. 4) and the total measured energy, $Q_{ac}$, (points in Fig. 4) would be lower than the predicted strain energy. Second, experiments are in contradiction with the hypothesis that $\Delta\theta$ increases with the defect concentration. Coffa et al. [41] determined, by electrical conductivity, the density of defects near the Fermi level and, by Raman spectroscopy, the bond angle dispersion on the same samples [12]. The results are summarized in the inset of Fig. 4. Although both parameters diminish with the annealing temperature, they do not keep any linear relationship. In addition we think that the results by Stolk et al. [24], already commented in Section III.2, are even more definitive because they show that no correlation exists between the TO Raman peak width and the defect concentration.

## IV. Conclusions

Our critical analysis of the theoretical relationships between the bond-angle dispersion and the Raman TO band width has shown that, once these magnitudes are quantified with the same parameters, the discrepancies between different authors are drastically reduced. In fact, the proposed relationships can be grouped either around the relationship of Vink et al. [10] or around that of Beeman et al. [9]. Experimental results show a slightly better agreement with the latter relationship. However, in view of the large error bars of the experimental points, it is clear that the use of any relationship for calculating the absolute value of $\Delta\theta$ should be done with caution. We consider that the information critically summarized in Fig. 3 gives a good state of the art on the subject.

In a second stage, the Raman TO half-width has been used to quantify the strain energy in pure a-Si. It has been concluded that in a-Si obtained by ion implantation, the heat evolved during relaxation is mainly due to the diminution of bond-angle strain, whereas the contribution of defect annealing is much smaller. This conclusion refers to the heat of relaxation and does not imply that defects play a secondary role in the relaxation kinetics. Probably, as supposed by Stolk et al. [24] and indicated by molecular dynamics simulation [27], the network rearrangement leading to lower strain energy is mediated by defect diffusion and annihilation. Concerning the contribution of defects to the heat of crystallization, it is very similar to that of bond-angle strain. Crystallization energy around 7 kJ/mol is predicted for defect-free, relaxed a-Si.

**Acknowledgement**

This work has been partially supported by the Spanish Programa Nacional de Materiales under contract number MAT2006-11144.

**Appendix: Relation between $U_{ac}$ and $Q_{ac}$**

In this appendix we will show that the heat evolved during relaxation and crystallization is nearly equal to the energy difference between the partially-relaxed amorphous state and the crystalline state. The energy difference at 0 K will be simply the heat evolved during the ideal process sketched in Fig. 5. First, the sample is heated until it reaches the relaxed state. The heat exchanged during this step is:

$$Q_1 = Q_{relax} - \int_0^{T_{max} \approx 1000K} mc^a \, dT , \qquad (A1)$$

where m is the sample mass, $c^a$ is the specific heat of a-Si and $Q_i > 0$ for exothermic processes. The value of $T_{max}$ can be taken as the maximum temperature of the calorimetric signal (see the inset of Fig. 3). Second, at $T_{MAX}$ the material crystallizes and releases the heat of crystallization:

$$Q_2 = Q_{cryst}. \tag{A2}$$

Finally, the heat related to cooling the crystallized state down to 0 K amounts to:

$$Q_3 = \int_0^{T_{MAX}} mc^c \, dT. \tag{A3}$$

Collecting the three terms we get:

$$U_{ac}(0K) = Q_{ac} - \int_0^{T_{MAX}} m\Delta c^{ac} dT, \tag{A4}$$

where $Q_{ac} \equiv Q_{relax} + Q_{cryst}$ and $\Delta c^{ac} \equiv c^a - c^c$. Finally, as we are interested in the static energy, the zero point energy (ZPE) must be subtracted from $U_{ac}(0\ K)$, i.e.:

$$U_{ac} = Q_{ac} - \int_0^{T_{MAX}} m\Delta c^{ac} \, dT - U_{ac}^{ZPE}. \tag{A5}$$

This formula allows us to calculate the correction term between $U_{ac}$ and $Q_{ac}$. $U_{ac}^{ZPE}$ can be calculated from the experimental density of vibrational states, $g_i(\omega)$, of a-Si and c-Si [42]:

$$U_{ac}^{ZPE} = V \int_0^{\omega_{MAX}} (\frac{1}{2}\hbar\omega)[g_a(\omega) - g_c(\omega)] d\omega, \tag{A6}$$

where V is the sample volume. We obtain $U_{ac}^{ZPE} \approx -0.5$ kJ/mol.

On the other hand, the specific heat term requires knowledge of $\Delta c^{ac}$. It has been measured for Ge [43] but not for Si:

$$\Delta c^{ac}(Ge) = 4.2\ 10^{-3}(T - 50)\ J/(mol\cdot K), \tag{A7}$$

where T is given in K. $\Delta c^{ac}$ has the same temperature dependence as the anharmonic contribution to the specific heat of c-Ge [44]:

$$\Delta c^h (c-Ge) = 2.0 \ 10^{-3} (T-50) \ J/(mol \cdot K), \tag{A8}$$

which indicates that the anharmonic effects are responsible for most of the specific heat excess of a-Ge with respect to c-Ge. In fact, comparing eqs. (A7) and (A8) leads to:

$$\Delta c^{ac} \approx 2\Delta c^h. \tag{A9}$$

This expression will also be considered valid for silicon, and, since $\Delta c^h$ (Si) has been measured [45], it allows $\Delta c^{ac}$ (Si) to be calculated:

$$\Delta c^{ac} (Si) \approx 5 \ 10^{-3} (T-200) \ J/(mol \cdot K). \tag{A10}$$

Integration of $\Delta c^{ac}$ (Si) from 0 to 1000 K gives 1.5 kJ/mol, which is similar to the result we would obtain with the estimation for $\Delta c^{ac}$ (Si) used by other authors [46] following different assumptions. Now, adding the $\Delta c^{ac}$ and $U_{ac}^{ZPE}$ terms leads to the final result:

$$U_{ac} \approx Q_{ac} - 1.0 \ kJ/mol. \tag{A11}$$

**Figure captions.-**

**Figure 1.-** Theoretical predictions (lines) of the relationship between the RMS bond angle dispersion relative to the tetrahedral angle and the half-width at high energy of the TO Raman peak. Solid symbols are experimental results [the $(\Gamma/2)$ and $\Delta\theta$ values of the diamond were taken from different sources]. The original relationships labeled "Wong" and "Maley", and the open symbols used different definitions for $\Delta\theta$ and $(\Gamma/2)$.

**Figure 2.-** Definition of the parameters used to quantify the width of the TO band. Inset: linear relationship between $(\Gamma/2)$ and FWHM/2 obtained from experimental Raman spectra.

**Figure 3.-** Heat released when partially-relaxed ion-implanted a-Si samples are transformed into crystalline silicon, $Q_{ac}$. The annealing temperature is detailed for several samples. Squares and stars refer to different authors. The sample with $\Delta\theta = 7.7°$ is crystallized without any significant relaxation. Inset: thermogram measured for the sample previously annealed at 573 K to show the meaning of the experimental components of $Q_{ac}$ [25].

**Figure 4.-** Alternative analysis of the experimental points of Fig. 3 supposing that the heat of relaxation arises exclusively from defect annihilation and that the heat of crystallization is due exclusively to strain. We argue that this analysis is not correct (see text). Inset: no linear relationship between $\Delta\theta$ and the density of defects near the Fermi level, $g(E_F)$, can be deduced from experiment [12, 41].

**Figure 5.-** Scheme that allows the energy difference to be calculated between the amorphous and crystalline state at 0 K, $U_{ac}(0\,K)$, through the heat exchanged during the process of heating and cooling down the sample.

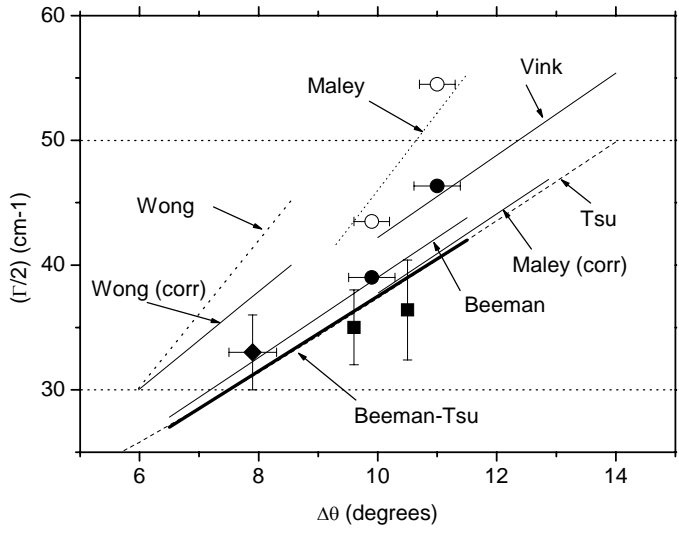

Figure 1.-

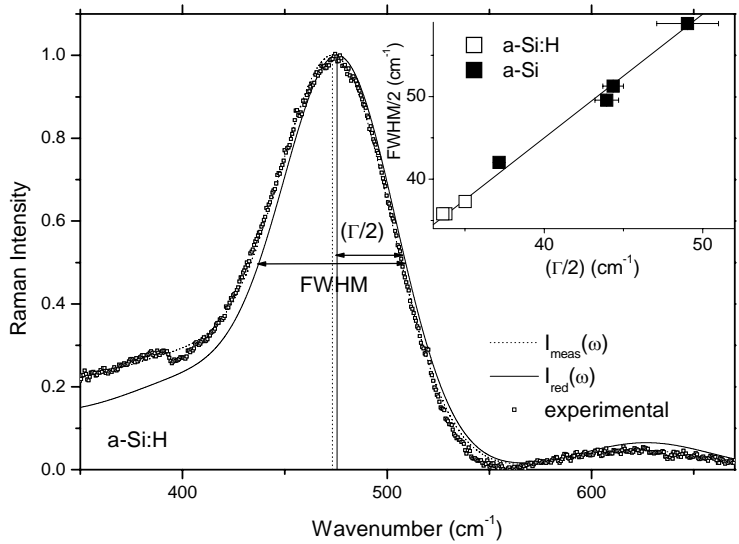

Figure 2.-

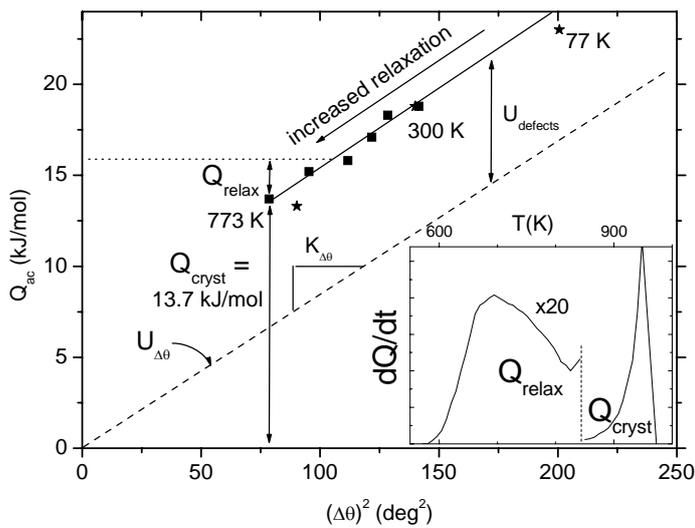

Figure 3.-

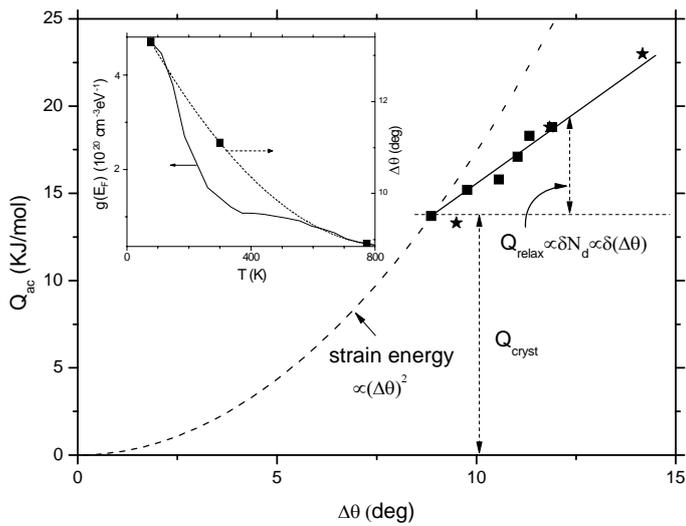

Figure 4.-

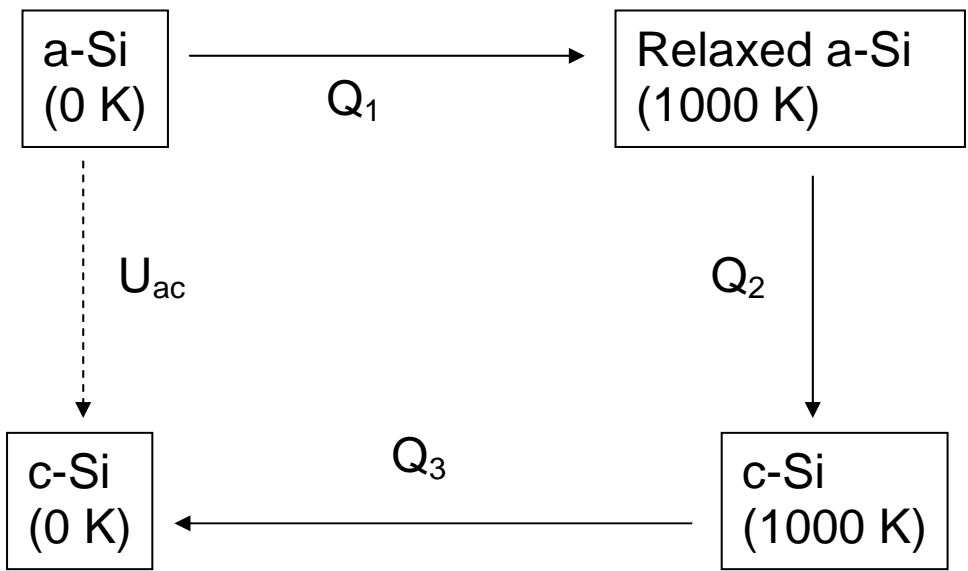

Figure 5.-